# The emojification of sentiment on social media: Collection and analysis of a longitudinal Twitter sentiment dataset

Wenjie Yin, Rabab Alkhalifa, Arkaitz Zubiaga
Queen Mary University of London, UK
{w.yin,r.a.a.alkhalifa,a.zubiaga}@qmul.ac.uk

**Abstract**

Social media, as a means for computer-mediated communication, has been extensively used to study the sentiment expressed by users around events or topics. There is however a gap in the longitudinal study of how sentiment evolved in social media over the years. To fill this gap, we develop TM-Senti, a new large-scale, distantly supervised Twitter sentiment dataset with over 184 million tweets and covering a time period of over seven years. We describe and assess our methodology to put together a large-scale, emoticon- and emoji-based labelled sentiment analysis dataset, along with an analysis of the resulting dataset. Our analysis highlights interesting temporal changes, among others in the increasing use of emojis over emoticons. We publicly release the dataset for further research in tasks including sentiment analysis and text classification of tweets. The dataset can be fully rehydrated including tweet metadata and without missing tweets thanks to the archive of tweets publicly available on the Internet Archive, which the dataset is based on.

## 1 Introduction

Social media research (Ngai, Tao, & Moon, 2015) and sentiment analysis (Poria, Hazarika, Majumder, & Mihalcea, 2020) have gained popularity in the last decade, both separately and jointly. Sentiment analysis of social media content is a fertile research area in a wide variety of fields including natural language processing (Wilson, Wiebe, & Hoffmann, 2005), computational social science (Vydiswaran et al., 2018), information science (Zimbra, Ghiassi, & Lee, 2016) and beyond.

As is the case for other similar problems in social media research, availability of datasets is key for any research project focusing on sentiment analysis. A good number of annotated sentiment datasets exists, among others the datasets published as part of SemEval shared tasks (Rosenthal, Farra,

& Nakov, 2017), but also in other languages including Arabic (Nabil, Aly, & Atiya, 2015), Russian (Rogers et al., 2018) or Norwegian (Øvrelid, Mæhlum, Barnes, & Velldal, 2020), for example. However these datasets are generally limited either in terms of size (Saif, Fernandez, He, & Alani, 2013; Nabil et al., 2015), in that they only provide tweet IDs with limited possibility of rehydration (Zubiaga, 2018) or in that they tend to cover a short period of time. Here we create and release TM-Senti, which aims to address these gaps by providing a large-scale, longitudinal and multilingual sentiment dataset, which can be fully rehydrated. This is thanks to the dataset being based on tweets publicly available on the Internet Archive. By releasing a collection of tweet IDs and their associated labels, researchers can regenerate the dataset with the same set of tweets and the same metadata. This enables creating a sentiment dataset that can be used for benchmarking in research.

In this paper, we make the following contributions:

- We curate and assess a list of emojis and emoticons that enable collection of large-scale sentiment datasets through distant supervision.
- To the best of our knowledge, our proposed methodology and dataset is the first to bring together longitudinal and multilingual dimensions to a sentiment dataset of this scale including 184 million posts.
- We conduct an analysis of the resulting dataset, showing among others that emojis are increasingly taking over the place of emoticons to express sentiment, with exceptions such as the sad face smiley that is still popular.
- We make all relevant resources available to the research community, including the TM-Senti dataset[1] as well as the code[2] enabling creation of new datasets following the same methodology.

Our work provides the first-of-its-kind resource and methodology for longitudinal sentiment analysis, with numerous applications ranging from longitudinal analyses of the evolution of sentiment across different topics, as well as the development of sentiment analysis methodologies with an awareness of changes in time. Among other findings, our analysis quantifies the extent of the emojification of sentiment in social media where emojis are increasingly prevailing over emoticons.

The rest of the paper is structured as follows: Next, we describe and provide background on distant supervision as a methodology to collect sentiment datasets. Then, we describe the methodology we develop for data collection and labelling. Then, we show the analysis of the dataset. Finally, we conclude the paper, discussing potential applications and usages of our work, as well as its limitations.

[1] `https://figshare.com/articles/dataset/TM-Senti/16438281`

[2] `https://github.com/AkaneNyan/distant-supervision-tweets`

# 2 Distant supervision

Distant supervision consists in compiling a list of evidence keywords (emojis, emoticons, hashtags) that can be used as stand-alone labels or aggregated representations for higher level labels to automatically annotate a target dataset. It has been extensively used especially for sentiment analysis (Go, Bhayani, & Huang, 2009), where one can use symbols like (':)') and (':(') as signals indicating that the text next to the symbol expresses a positive or negative sentiment, respectively. For example, given the following text:

*`I've enjoyed watching the show today :)`*

One can then assume that the text "I've enjoyed watching the show today", after removing the symbol, can be considered as an example expressing positive sentiment. Coming up with an extensive list of symbols for positive and negative sentiment then enables automated collection of texts labelled for sentiment.

A body of work has leveraged distant supervision for collecting datasets labelled for sentiment (Go et al., 2009), emotion (S. Mohammad, 2012) and stance detection (Kumar, 2018; S. M. Mohammad, Sobhani, & Kiritchenko, 2017), among others.

In this work, we curate an extensive list of emoticons and emojis that enable collection of data labelled for positive or negative sentiment, which we leverage for collection of a longitudinal sentiment dataset. Our interest to initiate this work was motivated by the lack of a dataset that would cover a longitudinal time period.

# 3 Data Collection and Labelling

In this section we describe the creation process of the dataset, which was carried out in three main steps: **(1)** data collection, **(2)** data labelling, **(3)** data sampling and **(4)** deduplication of tweets.

## 3.1 Data collection

We downloaded all the tweets dating from January 2013 to June 2020 which are available through the Twitter Stream Grab (TSG) project on the Internet Archive.[3] This is an archived collection of Twitter's 1% public stream of tweets, which amounts to 3.8TB worth of tweets in compressed format (bz2).

## 3.2 Data labelling.

We use distant supervision for sampling and labelling tweets from the large data collection retrieved from the TSG project. While a list of keywords had been defined by (Go et al., 2009), this has become obsolete especially due to the emergence of emojis. Therefore, we opted for expanding the list of keywords to includes both emoticons and emojis.

[3] `https://archive.org/details/twitterstream`

#### 3.2.1 Emoticons

For emoticons, we started with an extension to (Go et al., 2009)'s list: emoticons used for training sentiment embeddings by (Byrkjeland, de Lichtenberg, & Gambäck, 2018). With this initial list, we then needed to assess whether all of them were suitable for distant supervision. For this assessment, we chose to use manually labelled datasets, to assess the extent to which our lists of symbols matched a manually annotated dataset. Based on the Twitter sentiment datasets released as part of SemEval (Rosenthal et al., 2017) between 2013 and 2017, we tweaked the emoticon list by excluding semantically ambiguous terms.

The final evaluation matching *SemEval* tweets based on the predicted emotion by the distantly supervised approach in comparison with the manually annotated labels is shown in Table 1. While the manual labels are three-way (positive, neutral, negative), the distantly supervised approach only captures two different categories (positive and negative). This shows the difference in the inclusion thresholds used – the manually labelled "neutral" instances that our distant supervision captured with polarity mainly consists of mildly positive or mildly negative ones.

Considering the rest of the instances that were captured, little were noise (coloured red), showing the effectiveness of the distantly supervised method to automatically label tweets with little noise. These noise instances have similar distributions of emoticons as the valid ones, showing ironic use. Users were more likely to ironically use positive emoticons in a negative setting than vice versa.

| | **positive** | **neutral** | **negative** |
|---|---|---|---|
| **positive** | 510 | 95 | 19 |
| **negative** | 9 | 27 | 71 |

Table 1: Manual labels (columns) vs distantly supervised labels (rows) on the subset of SemEval tweets matching emoticons or emojis.

#### 3.2.2 Emojis

For emojis, we constructed our own list.

Starting from the list of emojis available at the point of September 2020, we first applied an initial filter by category, only to include emojis that are full faces or full gestures, excluding those that contain a body, clothing, or hairstyle. This was done to minimise the possible confounding effect of gender and culture.

For the remaining emojis after this initial filter, we considered the suitability of each emoji. To ensure we have an accurate interpretation of each emoji, we based our understanding on the text descriptions on Emojipedia.[4] We applied the following inclusion criteria on the descriptions: explicitly showing a named positive type of emotion, openness, or approval (positive); or explicitly showing a named negative emotion, rejection, or disapproval (negative). We covered strong and simple emotions (e.g. anger, happiness) as well as mild and complex ones (e.g. awkwardness, embarrassment). This was to increase the diversity of sentiment expressions in the dataset in terms of intensity and complexity.

Then, we further excluded emojis that contained external factors such as temperature and illness to further decrease possible ambiguity in the intention of the user.

Allowing our distantly supervised approach to cover a total of 137 balanced sentiment keywords: 69 positive and 68 negative expressions. The final list is shown in Appendix A: (1) Emoticons including 40 positive and 35 negative expressions, and (2) Emojis including 29 positive and 33 negative expressions.

### 3.3 Data sampling.

We sampled all the tweets that match more than one symbol of only one category. Specifically, we tokenised each tweet using white space and new line characters, and a match is where a positive or negative symbol was present in the tokens. Tweets matching both positive and negative symbols (which were very rare) were omitted.

We sampled tweets for seven different languages: Arabic (ar), German (de), English (en), Spanish (es), French (fr), Italian (it) and Chinese (zh).

### 3.4 Deduplication of tweets.

The sampled dataset contains both tweets and retweets, some of which creating duplicates in terms of textual content. In cases where both the original tweet and the retweets were present, we remove duplicates by keeping only the original tweet rather than the retweets. In cases where only retweets but not the original tweet were present, we deduplicate by only keeping the retweet that happened earliest in time.

## 4 Analysis of the Dataset

The final dataset contains 184,485,919 tweets, of which 108,248,874 (58.7%) are labelled as positive and 76,237,045 (41.3%) are labelled as negative. Next we analyse some of the key characteristics of this dataset.

[4] `https://emojipedia.org/`

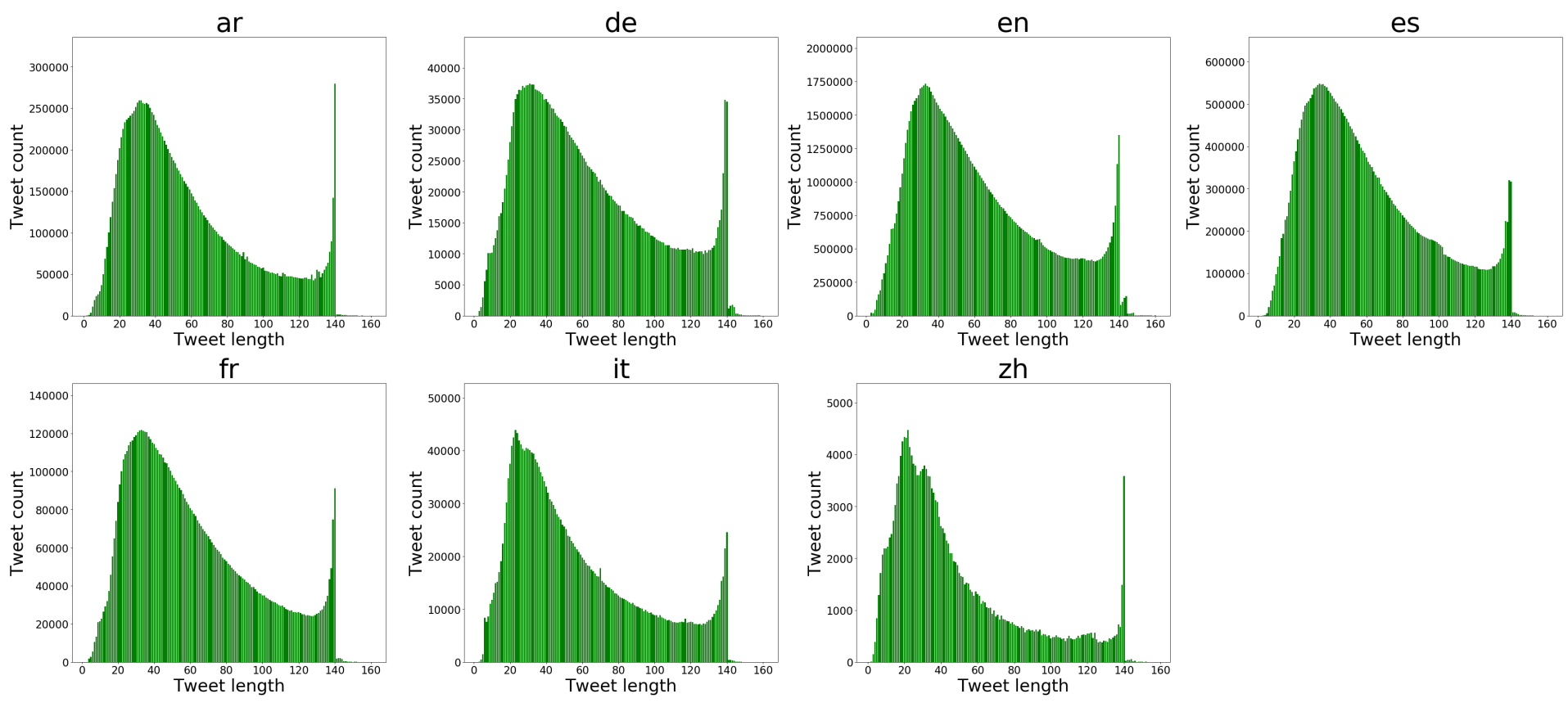

Figure 1: Tweet count by length.

**Users.** Tweets in the dataset were posted by 55,350,854 distinct authors. The most active Twitter user posted 7,717 tweets; total of 22,381,350 (40.4%) users posted more than one tweet, while the reminding 32,969,504 (59.6%) users posted only one tweet.

**Sources.** Tweets were posted using 47,678 different tools. Of these, the three most popular tools include: Twitter for iPhone with 88,801,158 (48.1%) tweets, Twitter for Android with 45,096,799 (24.4%) tweets and Twitter Web Client with 9,303,729 (5.0%) tweets.

**Tweet lengths.** Figure 1 shows tweet counts by length of tweet in characters. Trends are consistent for all languages, showing a cluster of tweets with lengths around 30 characters, a decreasing trend for longer tweets and a sudden spike for tweets around 140 characters.

**Temporal distribution of tweets.** Table 2 shows the number of tweets by year for the 7 languages. While there is an overall decreasing tendency in the number of tweets for most languages, Chinese is the exception exhibiting an increasing tendency. We believe this may be due to possible changes in the sampling strategy used by Twitter when providing the 1% tweet stream.

**Emoji vs emoticon usage.** Figure 2 shows the percentage of tweets that contain emoticons or emojis over time. All languages show a clear increasing tendency in the use of emojis over emoticons. While in 2013 there were languages were emoticons were more frequently used (i.e. Spanish, French and Italian), in the most recent data from 2020 all languages overwhelmingly use emojis over emoticons. This shows an interesting shift in the use of different social media conventions to express sentiment over time – with the constant update of new emojis and their popularisation, the use of emoticons has become increasingly rare.

**Top emojis and emoticons over time.** We also look at the top symbols used by year, to analyse how usage of the most popular emojis and emoti-cons has evolved over time. Table 3 shows the top 5 positive and top 5 negative symbols used each year between 2013 and 2020, aggregated for all 7 languages under study.

| Year | # Tweets | | | | | | |
|---|---|---|---|---|---|---|---|
| | **ar** | **de** | **en** | **es** | **fr** | **it** | **zh** |
| **2013** | 2,302,568 | 1,067,598 | 27,914,076 | 10,405,718 | 1,952,223 | 8,32,565 | 18,654 |
| **2014** | 2,495,167 | 3,74,272 | 28,811,423 | 7,054,278 | 1,622,567 | 5,30,925 | 25,620 |
| **2015** | 1,788,116 | 2,76,014 | 15,227,924 | 4,785,823 | 9,72,969 | 2,44,694 | 18,009 |
| **2016** | 2,329,499 | 3,18,878 | 14,569,976 | 5,294,192 | 9,85,689 | 2,50,190 | 19,348 |
| **2017** | 2,086,165 | 2,64,515 | 10,416,039 | 3,563,579 | 8,27,192 | 1,92,811 | 26,396 |
| **2018** | 1,660,273 | 1,82,602 | 6,979,167 | 2,223,653 | 6,12,629 | 1,39,473 | 26,356 |
| **2019** | 1,822,357 | 1,74,502 | 7,176,386 | 2,101,522 | 6,76,336 | 1,46,572 | 32,593 |
| **2020** | 1,587,829 | 1,45,066 | 6,311,522 | 1,867,447 | 5,85,168 | 1,30,163 | 38,631 |

Table 2: Tweets by year for all 7 languages.

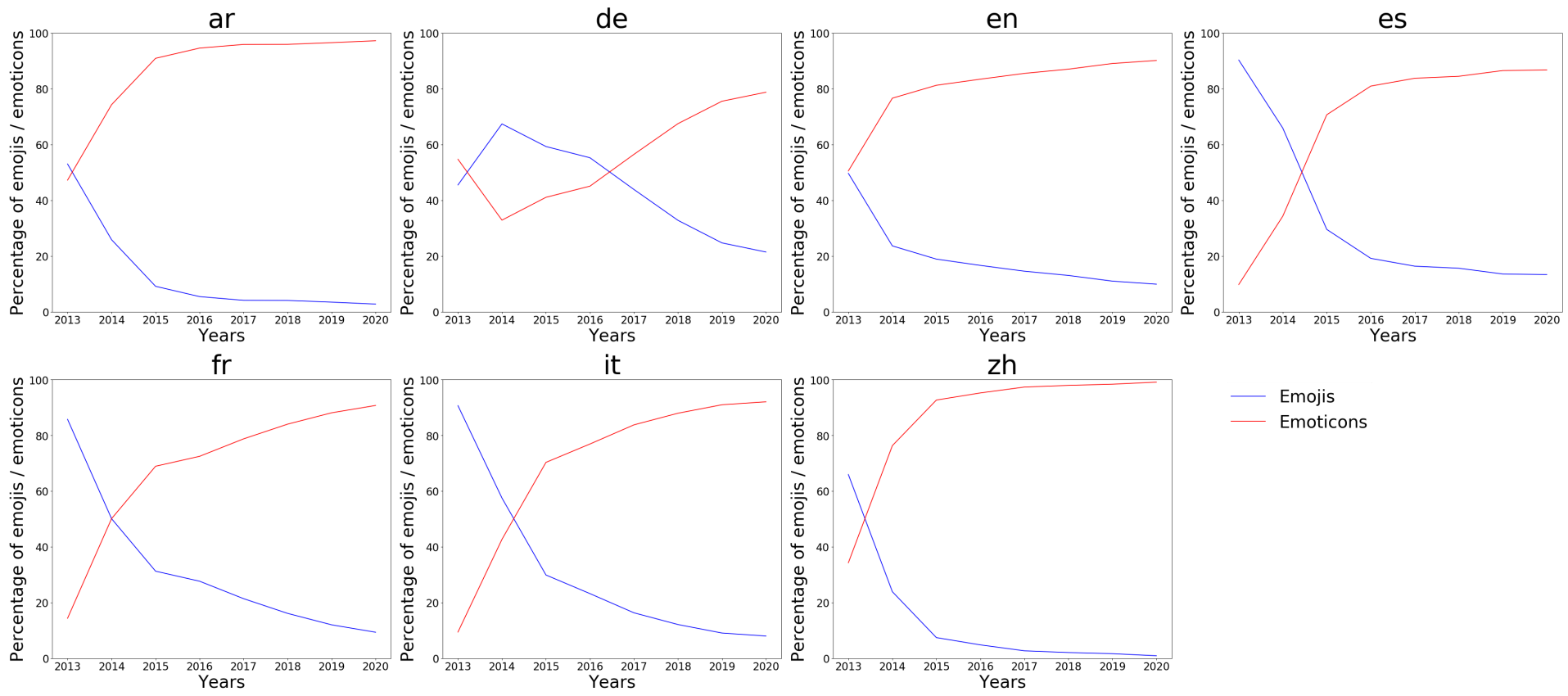

Figure 2: Emoji vs emoticon usage.

This again confirms that the popularity of emojis has increased over time at the expense of emoticons. Interestingly, this differs by sentiment polarity. Positive tweets started off largely dominated by emoticons in 2013. However, with the most positive emoticon (':)') dropping out from the top 5 list in 2019, top-5 popular symbols no longer contained any emoticons. In contrast, this decreasing use of emoticons is less drastic: for negative tweets, the (':(') emoticon remains on the top 2 position.

The relative popularity of different emojis stayed rather consistent, with the exceptions of a few emojis quickly gaining popularity soon after their introduction. We observe that the first most popular emoji remains consistent since 2015 for both the positive (😍) and negative (😭) classes. Top positions for the positive class are mostly consistent when it comes to the emojis, with the biggest change in 2020 as the new emoji (🥰) made it to the second position. When it comes to the negative emojis, the pensive face (😔) has increased in popularity in 2019 and 2020, whereas the weary face (😩) has gradually dropped in popularity from the third to the fifth position.

| Positive | | | | | | | |
|---|---|---|---|---|---|---|---|
| **2013** | **2014** | **2015** | **2016** | **2017** | **2018** | **2019** | **2020** |
| :) | :) | 😍 | 😍 | 😍 | 😍 | 😍 | 😍 |
| :D | 😍 | :) | :) | 😊 | 😊 | 😊 | 🥰 |
| ;) | 😘 | 😊 | 😊 | :) | 😘 | 👍 | 😌 |
| 😍 | 😊 | 😘 | 😘 | 😘 | :) | 😘 | 👍 |
| 😘 | 😌 | 😌 | 😌 | 😌 | 👍 | 😌 | 😊 |

| Negative | | | | | | | |
|---|---|---|---|---|---|---|---|
| **2013** | **2014** | **2015** | **2016** | **2017** | **2018** | **2019** | **2020** |
| :( | 😭 | 😭 | 😭 | 😭 | 😭 | 😭 | 😭 |
| 😭 | :( | :( | :( | :( | :( | :( | :( |
| 😩 | 😩 | 😩 | 🙄 | 🙄 | 🙄 | 😔 | 😔 |
| 😒 | 😒 | 😒 | 😩 | 😩 | 😩 | 🙄 | 🙄 |
| :/ | 😔 | 😔 | 😒 | 😢 | 😔 | 😩 | 😩 |

Table 3: Top positive and negative symbols (emojis and emoticons) by year, aggregated for all 7 languages.

**Temporal distribution of tweets by sentiment polarity.** Figure 3 shows the proportion of positive and negative tweets. With slight variations across languages, the figures are quite consistent in showing an increasing tendency of negative tweets, with the exception of Chinese, which plateaued in 2016 and then decreased. We interpret this overall decrease of positive tweets as a by-product of the decreasing use of emoticons mentioned above, which reduced the number of tweets captured by distant supervision overall. This effect is thus more significant in positive tweets, which showed a more dramatic drop in the use of emoticons.

**iPhone vs Android users.** Further digging into the differences in the use of emojis and emoticons, we look at the differences between iPhone and Android users. Table 4 shows the top positive and negative symbols for users of the two platforms. With both platforms using the same emoji as their top choice, (😍) for positive and (😭) for negative, we observe other interesting differences.

Overall, in both types of sentiment, Android users seem to use emoticons more often than iPhone users. We believe this is owing to the fact that emojis were introduced much later on Android devices (late 2012) than iPhone (2008). Furthermore, the specific choices of emojis within each type vary. For instance, if we look at the negative emojis, the crying face (😢)

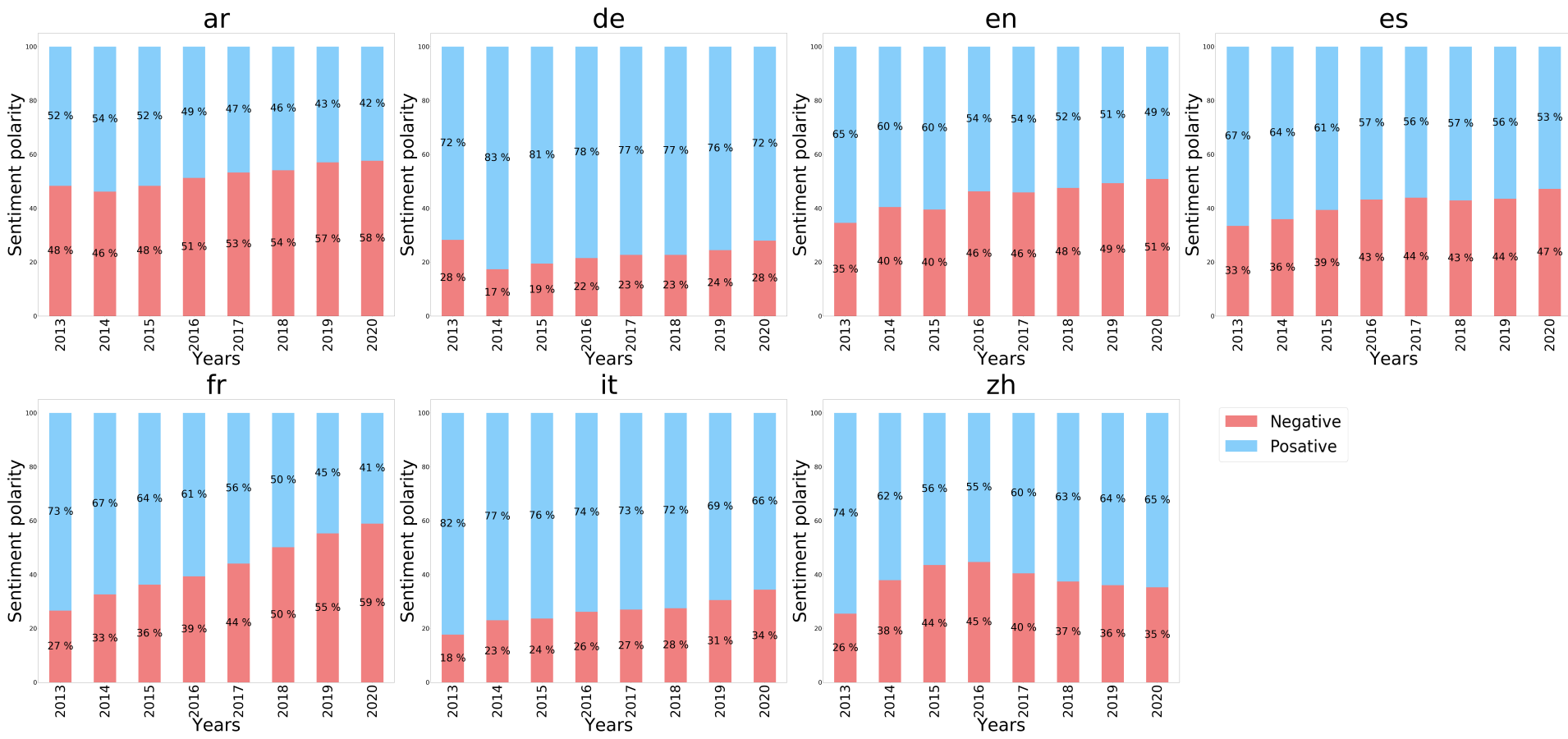

Figure 3: Yearly tweets by sentiment polarity, indicated as the ratio of negative vs positive tweets.

is commonly used by Android users whereas the weary face (😩) is more commonly used by iPhone users.

# 5 Discussion and Conclusion

Our work provides a methodology to create large-scale, longitudinal and multilingual sentiment datasets that can be used in a any scenarios and topics. Through this methodology, we collect a large-scale Twitter sentiment dataset in 7 languages, which can be rehydrated by researchers thanks to being sampled from a publicly available archive of tweets from the Internet Archive. This rehydration can be done through the tweet IDs we provide with our dataset and by sampling the tweets from the archive.

Thanks to the dataset, we performed a longitudinal analysis and observed a gradual change in the relative popularity of symbols for sentiment expression over time. The most prominent aspects include the preference for emojis over emoticons and the top symbols used. Specifically, users increasingly preferred emojis over emoticons throughout the period that we studied. Meanwhile, these trends noticeably differ by the type of sentiment polarity, with the change more drastic in positive tweets. Other factors also play a part, such as the platform on and language in which the tweets were posted, with the influence were rather limited compared to the temporal shift.

| Positive | |
|---|---|
| **Android** | **iPhone** |
| 😍 | 😍 |
| :) | 😘 |
| 😊 | 😊 |
| 😘 | 😌 |
| 😁 | :) |
| **Negative** | |
| **Android** | **iPhone** |
| 😭 | 😭 |
| :( | 😩 |
| 😒 | :( |
| 😢 | 😒 |
| 😔 | 😔 |

Table 4: Top positive and negative symbols used by Android vs iPhone users.

### 5.1 Applications and Usage

Our methodology to develop longitudinal sentiment datasets enables further research in a number of directions, some of which we discuss next:

- Longitudinal analyses of sentiment can be valuable for longstanding topics or events, however labelling data can be unaffordable if done through manual labelling. Our methodology can enable the longitudinal analysis of events, for example studying how sentiment has changed over events such as Brexit (Hürlimann et al., 2016).
- Recent research is showing that machine learning models for classification of social media posts fade over time in terms of performance (Florio, Basile, Polignano, Basile, & Patti, 2020; Alkhalifa, Kochkina, & Zubiaga, 2021). Where one uses labelled data from a particular point in time for training machine learning models, performance of these models drops when they are applied on temporally distant data, among others because language and communication in social media changes over time and models need to be adapted to capture that change. Availability of longitudinally labelled datasets can help further study this problem in the context of sentiment analysis.
- Given that our methodology is not restricted to a specific language, it can be applied to a wide range of languages to enable multilingual analyses of social media, as well as development of multi- (Dashtipour et al., 2016) and cross-lingual (Zhou, Wan, & Xiao, 2016) models for sentiment analysis. While we haven't tested our methodology on other social media platforms, we anticipate that our approach can be effectively applied to other platforms to enable broader sentiment analysis.

### 5.2 Limitations

While the methodology and analyses in this paper provides the means for studying and developing sentiment analysis techniques at scale, it comes with a number of limitations that need to be taken into consideration:

- We rely on the TSG project to generate our dataset. Given that the data available through this project is restricted to the 1% sample available through the free Twitter API, our dataset inevitably has the same restrictions (Morstatter, Pfeffer, Liu, & Carley, 2013). Our methodology and code is however flexible in that it can be applied to any dataset, for example restricted collections of tweets linked to a particular topic.

- Tweets provided in our dataset are labelled through distant supervision, as has been validated in previous work, but being an imperfect method it also leads to some noisy labels. Instances where smileys are used ironically are not avoidable, although rather limited in number, as shown in our validation analysis.

- We have curated a list of emojis and emoticons from the pool of symbols available as of 2020. While new emojis continue to be created, the list of symbols may need revision to incorporate relevant ones.

## Acknowledgments

# A List of Emojis and Emoticons for Distant Supervision

| Emoticons | | | | Emojis | | | |
|---|---|---|---|---|---|---|---|
| **Positive** | | | | | | | |
| :) | :-) | ;D | :D | 😀 | 😃 | 😄 | 😁 |
| =D | :-] | :] | :-3 | 😆 | ☺️ | 😊 | 😇 |
| :3 | :-> | :> | 8-) | 😌 | 😍 | 🤩 | 🥰 |
| :-} | :} | :o) | :c) | 😘 | 😗 | 😚 | 😙 |
| :)̂ | =] | =) | :D | 😋 | 😝 | 🤗 | 😌 |
| 8-D | x-D | xD | X-D | 🤤 | 🥳 | 😎 | 😺 |
| XD | =D | =3 | B^D | 😸 | 😻 | 😽 | ✌️ |
| :-)) | :'-) | :') | ;-) | | 👍 | | |
| ;) | *-) | *) | ;-] | | | | |
| ;] | ;)̂ | :-, | <3 | | | | |
| **Negative** | | | | | | | |
| :( | :-( | : ( | :'( | 🙃 | 😐 | 😑 | 😶 |
| :L | =L | :c | :c | 😒 | 🙄 | 😬 | 😔 |
| :-< | :< | :-[ | :[ | 🤢 | 🤮 | 😵 | 😕 |
| :-\|\| | >:[ | :{ | :@ | 😟 | 🙁 | ☹️ | 😦 |
| >:( | D': | D:< | D: | 😧 | 😨 | 😰 | 😢 |
| D; | D= | :-/ | :/ | 😭 | 👎 | 😖 | 😣 |
| :-. | >:\ | >:/ | :\ | 😞 | 😓 | 😩 | 😫 |
| =/ | =\ | >.< | v.v | 😡 | 😠 | 🤬 | 😿 |
| :S | </3 | <\3 | | | 🖕 | | |

Table 5: Positive and negative sentiment keywords